\documentstyle[amsfonts,aps,multicol,tighten,epsfig]{revtex}
\begin{document}

\title{Entropy production in the cyclic lattice Lotka-Volterra model}
\author{Celia Anteneodo}    
\address{Centro Brasileiro de Pesquisas F\'{\i}sicas,\\             
Rua Dr. Xavier Sigaud 150, 22290-180,             
Rio de Janeiro, RJ, Brazil.}  

\maketitle    
\begin{abstract} 

The cyclic Lotka-Volterra model in a $D$-dimensional regular 
lattice is considered. Its ``nucleus growth'' mode is analyzed 
under the scope of Tsallis' entropies 
$S_q=(1-\sum_i p_i^q)/(q-1)$, $q\in \mathbb{R}$. It is shown both 
numerically and by means of analytical considerations that a 
linear increase of entropy with time, meaning finite asymptotic entropy rate, 
is achieved for the entropic index $q_c=1-1/D$. 
Although the lattice exhibits fractal patterns along its evolution, 
the characteristic value of $q$ can be interpreted in terms of 
very simple features of the dynamics. 

\end{abstract}  

\begin{multicols}{2} 
\narrowtext 

\section{Introduction}
Over a decade ago, Tsallis proposed an entropic form as a starting  
point for a possible 
generalization  of Boltzmann-Gibbs statistics \cite{ct88} 
(see also \cite{review} for a review on the subject).
The  generalized entropy has the form 

\begin{equation}
\label{Sq}
S_q \;=\; k\frac{1-{\displaystyle \int {\rm d}x\,
[\rho(x)]^q} }{q-1} \, ,\hspace*{1cm} \mbox{with $q\in\mathbb{R}$},
\end{equation}
where $k$ is a positive constant and  $\rho$ a normalized probability density. 
The usual Boltzmann-Gibbs-Shannon (BGS) entropy is recovered when $q=1$ 
[$S_1=-k\int{\rm d}x\rho(x) \ln \rho(x)$]. 

Many experimental and numerical data 
are well approximated by $q$-exponentials (the probability distributions 
that maximize $S_q$ under simple constraints), giving support to the 
applicability of the new entropic measure to those systems.  
In the case of the high-dimensional systems of interest in statistical 
physics, first-principle derivations are still to be made. 
However, Tsallis' entropy seems to be the appropriate one for 
certain low-dimensional dynamical systems, such as unimodal maps at the 
edge of chaos, where the usual Lyapunov exponent vanishes. 
In such cases the characteristic value of $q$ is related to 
relevant properties of the dynamics and it can be calculated 
by several paths, 
such as the multifractal spectrum of the attractor \cite{multifrac} or   
the sensitivity to initial conditions \cite{sic}, or even, 
from a rigorous renormalization group analysis \cite{rg}.  
Furthermore, this scenario has been shown to be valid also   
for two-dimensional maps \cite{twod}. 

One of the properties of  $S_q$ that, for an appropriate value of the 
entropic index, can make it in  some cases preferable to the standard 
entropy  is the possibility of having an asymptotic rate of entropy production with 
a non-trivial value. 
Among the {\em a priori} infinite possible values of $q$, 
the one leading to a linear increase of the entropy  with time, 
implying {\em finite} rate of entropy growth, is selected as being $q_c$, 
a value of $q$ characteristic of the system. 
In fact, this is another path for the determination of $q$ that in  the 
case of unimodal maps provides the same results \cite{baranger} as   
the alternative methods mentioned above. 
Moreover, rigorous analytical results have recently been found for such  
systems along this line \cite{pesin}. 
The temporal evolution of the $S_q$ family, allowing to determine $q_c$, 
has also been investigated for diverse 
other dynamical systems \cite{logistic,hyper,boghosian,tpt04}.     
The outcome $q_c \neq 1$ has usually been interpreted as a 
signal of nonextensivity. 

Under the scope of the generalized entropies  $S_q$, we will study 
in the present paper the temporal evolution of a cyclic Lotka-Volterra 
model in the lattice (LLV) \cite{tainaka,pnb99}. 
This model is particularly rich as it exhibits interesting features, 
such as stationary states with spatial patterns \cite{tainaka} 
and fractality \cite{fractals}, which makes it a potential 
candidate for the  applicability of the entropies $S_q$. The LLV is one of the 
descendants of the original versions constructed by Lotka\cite{lotka} 
and Volterra  \cite{volterra} to model autocatalytic chemical reactions 
and the prey-predator dynamics, respectively, and further 
adapted to many other situations from active transport 
by proteins \cite{bisch} to social processes \cite{epstein}. 
The generalized LLV extends the original scheme to $\cal N$ species  
$A_i$ ($i=1,\ldots,\cal N$) such that, cyclically, 
$A_i$ is ``predator'' of $A_{i-1}$ and simultaneously  ``prey'' of  
$A_{i+1}$ (with $A_{{\cal N}+1}\equiv A_1$). 
Furthermore, the dynamics takes place over a lattice, 
an ingredient that introduces new interesting spatial 
features in comparison to the spatially homogeneous mean-field 
description \cite{space}.  

The criterion of finite entropy rate  has been applied before to 
the LLV model in one and two dimensions \cite{tpt04} and the relation 
$q_c=1-1/D$ was conjectured for arbitrary $D$.    
Now we will extend that study to higher dimensions in order to 
test the conjecture. 
Moreover, we will essay an interpretation of the connection 
between $q_c$ and space dimensionality, showing that $q_c\neq 1$ is not 
necessarily related to the fractal properties of the dynamics. 
The remaining of the paper is organized as follows. In Sect. II  
we describe in detail the specific LLV model considered. 
In Sect. III  the  time behavior of  
the entropies $S_q$ associated to the LLV is studied. 
Sect. IV exhibits the behavior of $S_q$ under 
simple transformations of a probability density leading to reanalyze 
in Sect. V the LLV dynamics. 
Final remarks are presented  in Sect. VI.

\section{The generalized Lotka-Volterra model in a lattice}

Particles of the $\cal N$ different species $A_i$ 
are localized at the sites of a $D$-dimensional hypercubic lattice. 
Reactions between particles of different species occur in bimolecular 
autocatalytic steps following the scheme

\begin{equation} 
\label{scheme}
A_i + A_{i+1} \stackrel{k_i}{\longrightarrow}  2A_{i+1}, 
\end{equation}
for $i=1,\ldots,\cal N$,  being $A_{{\cal N}+1}\equiv A_1$, 
and where $0\le k_i\le 1$  are the kinetic rates. 
No sites are empty but one of the species could be 
interpreted as representing empty sites in the lattice 
\cite{tpt04,pnb99,fractals,pt03}. 
The dynamics is implemented by means of a Monte Carlo (MC) 
algorithm following the details in Refs. \cite{tpt04,pnb99,fractals}. 
Basically, at every  microscopic step: (i) one lattice site is randomly 
chosen; (ii) one of its nearest neighbors is randomly chosen; 
(iii) if the first site is $A_i$ and the neighbor $A_{i+1}$, 
the first site changes to $A_{i+1}$ with probability $k_i$, 
in accord with scheme  (\ref{scheme}), otherwise the site 
remains unchanged. Each MC step (MCS, our unit of time) consists 
in $N=L^D$ microscopic steps defined above, where $L$ is the 
linear size of the lattice, so that at each MCS all sites 
are revisited once in average. Boundary conditions are periodic. 
In this paper we will deal with three species only (${ \cal N} =3$) and 
we will focus on the symmetric case where  
all the kinetic rates are equal to one ($k_1=k_2=k_3=1$). 
Moreover, we will restrict our study  to a particular mode of the 
dynamics, the  ``nucleus growth'' one \cite{tpt04,pt03}.  
The lattice is set with all the sites filled with species $A_3$ 
and a ``droplet'' or nucleus represented by a sublattice of small linear size 
$\lambda$ (i.e., $\lambda \ll L$) is introduced. The droplet  
contains equal amounts of particles of all the three species 
randomly and uniformly distributed.

\begin{figure}[htb]  
\begin{center}
\includegraphics*[bb=96 220 477 680, width=0.4\textwidth]{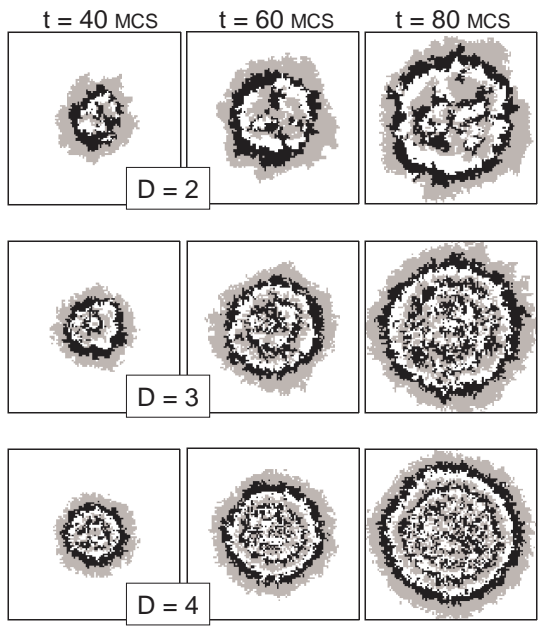} 
\end{center}
\caption{Snapshots of the dynamics at different times indicated on the top 
of the figure, for $D=2,\,3$ and 4. 
The initial condition is a droplet, where all the species are present in 
equal amounts and homogeneously distributed, over a background of $A_3$. 
In all cases the linear size of the lattice is $L=100$ and 
the initial nucleus has $\lambda=6$. 
For $D>2$ the snapshots correspond to sections parallel to one of the  
hypercube faces and passing by the position of the center of the 
initial droplet. 
Species are $A_1$ (gray), $A_2$ (black) and $A_3$ (white). 
}
\label{fig:fig1}
\end{figure}

As the system evolves according to the MC dynamics, the droplet 
grows and acquires spontaneously a peculiar structure \cite{poison}.   
Typical snapshots of the dynamics are exhibited in Fig.~\ref{fig:fig1} 
for lattices in $D=2,\,3$ and 4 dimensions. 
Rings of alternating species develop, 
repeating the sequence $A_1,A_2,A_3$  towards the center of the droplet.    
This pattern is possible since all the  kinetic constants are equal, 
then layers have almost the same radial velocity. 
The thickness of the rings decreases towards the center and  thickness fluctuations 
destroy the most immersed rings. This behavior has already been observed before 
by Provata  and Tsekouras for the two dimensional case \cite{pt03}. They also 
observed  that the destroyed rings give rise to a spatial organization of the 
species in  domains with fractal boundaries typical of the 
steady state in fully occupied periodical lattices\cite{tpt04,fractals,pt03}. 
As the dimensionality increases, spatial features remain qualitatively similar 
but length scales become shorter. 
In the mean-field limit $D\to\infty$, homogeneity is expected. 
We are going to inspect immediately the evolution of spatial patterns from 
the viewpoint of the generalized entropies $S_q$.

\section{Temporal evolution of $S_{\lowercase{q}}$ in the LLV}

Following previous work \cite{tpt04}, 
we study the temporal evolution of the entropies  
associated to one of the species, e.g., $A_1$. 
This particular choice does not substantially affect the results. 
Nonoverlapping windows or sublattices $\{W_i, \;1\le i\le M \}$ 
of edge length $l$ covering all the lattice are considered. 
Thus the number of windows must be  $M=(L/l)^D$, where $l$ is a divisor of $L$. 
We associate to each window $i$ a probability $p_i(t)$ 
of being occupied by species $A_1$ at time $t$ by  counting 
the number of particles of that species $n_1(i,t)$.   
Then $p_i(t)=n_1(i,t)/n_1(t)$, where $n_1(t)$ is the  
total number of particles $A_1$ on the lattice. 
The resulting set of probabilities is used to calculate 
the entropy of the lattice, given by the discrete version of 
Eq.~(\ref{Sq}) (where we have set $k=1$)

\begin{equation}       
\label{Sqdis}
S_q \;=\; \frac{1-{\displaystyle \sum_i p_i^q} }{q-1} \, .
\end{equation}

\begin{figure}[htb] 
\begin{center}
\includegraphics*[bb=106 244 520 640, width=0.5\textwidth]{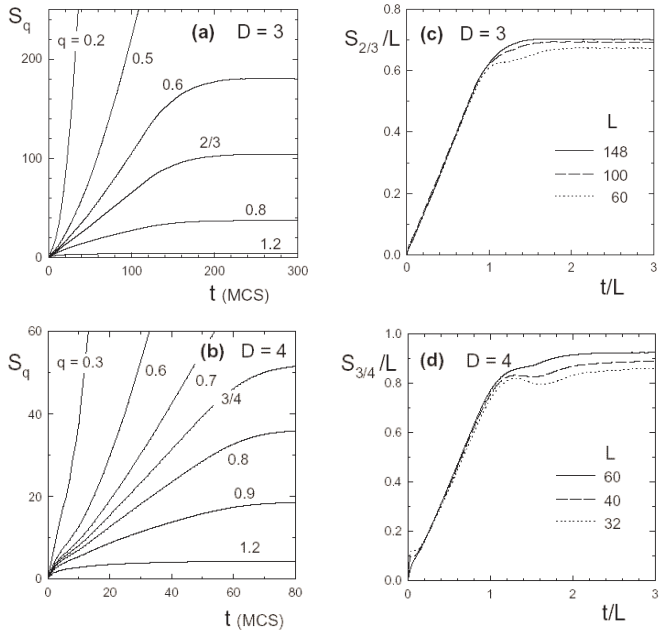} 
\end{center}
\caption{Time evolution of the generalized entropies. $S_q$ vs. $t$ for 
various values of $q$ and $L=148$ in  $D=3$ (a), $L=60$ in $D=4$ (b).
$S_q/L$ vs. $t/L$ for different lattice sizes $L$ with: $q=2/3$ 
when $D=3$ (c), 
$q=3/4$ when $D=4$ (d). Symbols correspond to a single representative 
numerical experiment. In all cases, the window of the partitioning 
is $l=4$ and the initial droplet size $\lambda=4$.}
\label{fig:sq}
\end{figure}

The LLV in $D=1$ and 2 dimensions has already been studied 
from this viewpoint before\cite{tpt04}. 
Now we will investigate higher dimensional  lattices. 
Fig.~\ref{fig:sq} shows the generalized entropies as a function of time 
for diverse values  of $q$,  when the lattice has $D=3$ and 4 dimensions. 
The results shown in Fig.~\ref{fig:sq} are not qualitatively affected by 
changing $l$ and $L$, as long as $1\le l \ll L$. 
They mainly affect the saturation level, 
$S_q^{sat.}=[(L/l)^{D(1-q)}-1]/(1-q)$ for a uniform distribution.

\begin{figure}[htb] 
\begin{center}
\includegraphics*[bb=128 326 483 600, width=0.3\textwidth]{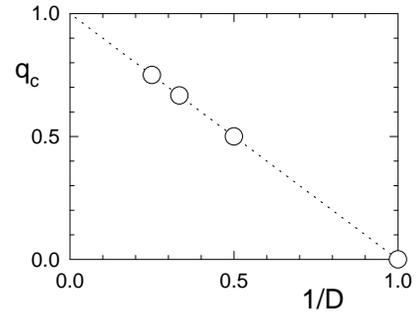} 
\end{center}
\caption{\protect 
Characteristic value $q_c$ as a function of the lattice dimensionality $D$. 
The dotted straight line corresponds  to $q_c=1-1/D$. 
Symbols are the result of numerical experiments. 
Errors are of the order of symbol size. For $D=3$ and 4:  
$q_c$ was numerically determined from the data in
Fig.~\protect\ref{fig:sq} as the value of $q$ 
yielding a linear slope. For $D=1$ and 2,  $q_c$ was extracted from 
Ref.~\protect\cite{tpt04}.  
}
\label{fig:qd}
\end{figure}

Notice in Figs. \ref{fig:sq}.a and \ref{fig:sq}.b that as $q$ increases the 
concavity passes from positive to negative, before saturation, 
which occurs when the droplet radius becomes of the order of the 
linear size of the lattice. 
Entropies with $q$ yielding constant slope (null concavity) 
are represented as a function of time in Figs. \ref{fig:sq}.c and \ref{fig:sq}.d. 
In all cases, constant slope occurs at a value of the entropic 
index $q_c$ that as function of $D$ follows the law

\begin{equation} 
\label{qc}
q_c\;\simeq\; 1-\frac{1}{D} \, ,
\end{equation}

\noindent
in agreement with the relation numerically found for $D=1$ and 2, and 
conjectured for generic $D$ previously \cite{tpt04}. 
See Fig.~\ref{fig:qd}.

\section{Behavior of $S_{\lowercase{q}}$ under simple 
transformations of a probability density}
\label{sec:cases}

The characteristic entropic index $q_c$ is related to the number 
of dimensions $D$ though a simple law.  
This leads to think that a simple mechanism could be behind. 
Since we are dealing with a growth process, we will investigate in this Section 
the relation between $q_c$ and $D$ resulting from some basic growth mechanisms  
and   connect  it  with Eq. (\ref{qc}). 

Let us analyze  the behavior of the entropies $S_q$ under   
a rescaling transformation of an arbitrary probability density $\rho(x)$,  
being $x$ a point in a $D$-dimensional space \cite{complex}.
The entropy of the rescaled function $\rho_\sigma(x)= \rho(x/\sigma )/\sigma^D$,  
where $\sigma$ is a linear length, becomes 

\begin{equation}  \label{scaling}
S_q(\sigma) \;=\; \frac{\sigma^{(1-q)D}\Bigl( (1-q)S_q(1) +1  \Bigr)-1}{1-q}.
\end{equation}
As $\sigma \to \infty$, the distribution broadens and the entropy grows     
due to the loss of order. 
If the scaling or stretching parameter increases exponentially with time, 
linear entropy increase occurs for $q=1$.  
If the scaling parameter follows the law 
$\sigma \sim t^{\gamma/2}$ with $\gamma>0$,  then 
the generalized entropies  increase  with time as follows: 
For $q<1$, they scale with time as $S_q(t)\sim t^{(1-q)D\gamma/2}$, 
$S_q(t)\sim \ln t$ in the marginal case $q=1$, and 
for $q>1$, saturation occurs  at  $S_q(\infty)=1/(q-1)$.  
Hence a linear increase is achieved in the regime $q<1$ for 

\begin{equation} 
\label{qdgamma}
q_c\;=\; 1-\frac{2}{\gamma D} \, .
\end{equation}

As a relevant particular case let us mention the normal diffusive 
spreading of a  $D$-dimensional Gaussian distribution where the 
stretching parameter is $\sigma^2(t)=2Qt$, with $Q$ the positive 
diffusion constant. 
In this case $\gamma=1$ then  $q_c=1-2/D$. The general case given 
by Eq.~(\ref{qdgamma}) may be due to anomalous diffusive motion. 
A still simpler example of rescaling transformation that will  
be useful later consists in a density being non-null and uniform 
inside a $D$-dimensional hypersphere and zero outside, such that 
the radius $\sigma(t)$ grows as $\sigma \sim t^{\gamma/2}$. 
Notice that the expression for $q_c$ of the 
growing LLV droplet, given by  Eq.~(\ref{qc}), 
is obtained when the temporal variation 
of the scaling length $\sigma$ is ballistic ($\gamma=2$).

Now, let us consider another simple process. 
$M \gg N$ windows cover a lattice, like for the LLV in the previous Section, 
with probabilities 
$$
(p_1,p_2,\ldots,p_N,\underbrace{0,0,\ldots,0}_{M-N}), \;\;\;\;\; \sum_{1\le i\le N} p_i=1.
$$ 
Imagine that windows with non-null probability are replicated by an integer factor $m$ 
such that the new set of probabilities is 
$$
(\underbrace{\frac{p_1}{m},\ldots ,\frac{p_1}{m}}_m,
\underbrace{\frac{p_2}{m},\ldots ,\frac{p_2}{m}}_m, \ldots,
\underbrace{\frac{p_N}{m},\ldots, \frac{p_N}{m}}_m,
\underbrace{0,0,\ldots,0}_{M-mN})
$$ 
If the replication factor is $m\sim\sigma^D$  (where $\sigma$ is a typical linear 
size of the $D$-dimensional system) and  additionally if $\sigma \sim t^{\gamma/2}$,  
then the entropy will be produced in such a way that a growth linear in time occurs 
for $q_c$ given by Eq. (\ref{qdgamma}). 
In fact, if a continuous distribution is discretized,  
the associated rescaling process 
can be thought as a particular  case  of the replication one here  considered, 
a special case where replicas are spatially ordered. 
Nevertheless for  discrete probabilities, the entropic form $S_q$ 
does not depend on the spatial localization of the windows.

Summarizing, while for an ordinary exponential growth process, 
the standard entropy  $S_1$ increases linearly with $t$;   
for a process where growth follows a power-law in time, 
$S_q$  has the property of finite asymptotic entropy rate  
for some $q=q_c\neq 1$. 
This value of the entropic index can result from simple 
transformations and therefore may be 
trivially related to the lattice dimensionality.   
These ideas lead us to review the results 
of the precedent Section to see whether simple mechanisms are also 
lurking there.

\section{Some details of the LLV dynamics}
\label{sec:details}

\begin{figure}[htb]  
\begin{center}
\includegraphics*[bb=60 130 526 690, width=0.35\textwidth]{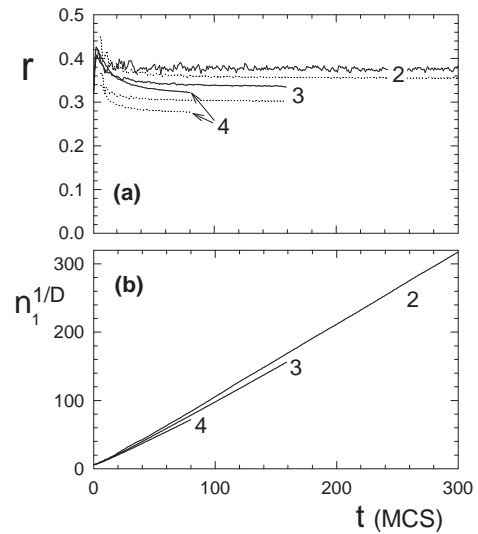} 
\end{center}
\caption{Propagation of a droplet of $A_1$ in a lattice filled with $A_3$,
for different values of $D$ indicated in  the figure.
(a) Reactivity $r$ as a function of time for the two-species LLV (full lines); 
for comparison, $r$ for a regularly filled hypersphere with the same total number 
of cells as the corresponding LLV nucleus is also plotted (dotted lines).  
(b) Total number of sites $n_1$ occupied by  $A_1$ vs. time.
The initial nucleus has linear size $\lambda=4$.}
\label{fig:rough}
\end{figure}

Let us inspect first the roughness of the interface because the reaction rate 
and therefore the dynamics of the propagating front  are  
related to that quantity. 
Since we are interested in the interface, 
one can perform numerical simulations considering just two species. 
Hence we follow the evolution of an initial nucleus containing only $A_1$ 
particles dropped over a background of $A_3$. Its propagation is 
like in the LLV case where a forefront of $A_1$ 
particles governs the  spreading of  the droplet. 
Since the reaction rate must be proportional to the ``extent'' of the interface, 
we measure $N_s$, the number of reactive sites constituting the interface, 
and $N_f$, the total number of reactive faces in interfacial sites. 
Reactive faces are those separating two nearest neighboring sites occupied by  
different species. A suitable quantity for our model is 

\begin{equation} \label{r}
r \;=\; \frac{N_f}{c N_s} \;,
\end{equation}
where $c$ is the connectivity, i.e, the number of  first nearest neighbors 
{\em per} site ($c=2D$ in hypercubic lattices). 
The quantity $r$ represents an averaged measure of the degree of 
reactivity of an interfacial site, and must be also connected to the 
roughness of the interface. 
The reactivity $r$ is plotted in Fig.~\ref{fig:rough}.a as a 
function of time.
It  soon  reaches a stationary value within small fluctuations.  
For comparison, the figure also exhibits the value of $r$ for a 
regularly filled hypersphere with the same total number of cells 
as in the LLV nucleus at each given  $t$. 
The higher the lattice dimensionality, the larger the relative difference 
between the values of $r$ for the two models. 
The roughness of the propagation front in the 1D dynamics has been studied 
in detail by Provata and Tsekouras \cite{pt03}. 
Also in this case, it was shown that after a brief transient the rough profile
remains stable in average.

If the reactivity remains constant in time,  the  front moves    
at constant  radial speed: $R(t)=R(0) + v\,t$, where $R$ is an effective 
radius $R\propto n_1^{1/D}$ and the  velocity $v\equiv v(r)$ an 
increasing function of its argument for a given $D$. 
Although the fractal properties of the interface are embodied in the velocity,  
if the roughness soon reaches its steady value, then fractality does not affect 
the temporal law of the front propagation. 
In effect, the linear dependence of $R$ with time is observed in 
numerical experiments (see Fig.~\ref{fig:rough}.b). 
Therefore, following the considerations of the precedent Section, 
Eq. (\ref{qdgamma}) with $\gamma=2$ must clearly hold in this binary case.

\begin{figure}[htb]  
\begin{center}
\includegraphics*[bb=70 220 540 740, width=0.35\textwidth]{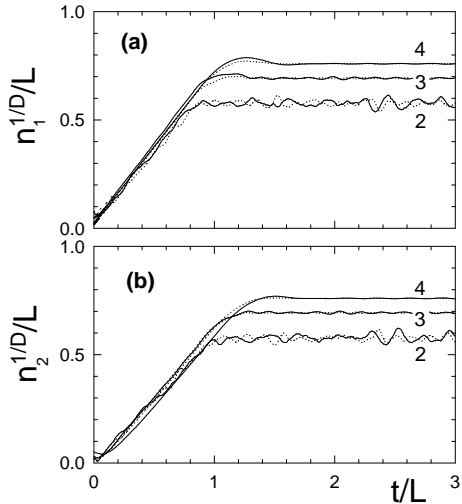} 
\end{center}
\caption{Time evolution of the number of cells $n_1$ and $n_2$ occupied by species 
$A_1$ (a) and $A_2$  (b),  respectively, 
for the values of the lattice dimensionality indicated in the 
figure and for various lattice sizes $L$: 150 and 200 ($D=2$), 
100 and 150 ($D=3$), and 60 and 100 ($D=4$), represented by  
dotted  and full lines respectively. 
Time is measured in MCS. Curves correspond to single runs. 
The initial droplet size is $\lambda=6$ for $D=2$ and $\lambda=4$ for $D=3,4$.}
\label{fig:n1n2}
\end{figure}

In the three-species case, although the reaction scheme is cyclic, 
symmetry is broken by the initial condition. 
Species $A_3$  plays the special role of a background what in turn 
makes $A_1$ play the special role of the forefront species.  
After a short transient and before the limits of the lattice 
are reached by the nucleus, the stationarity of the interface 
reactivity $r$, leads  to a linear growth of the effective radius with time, 
as in the two species case. 
Consequently, the total number of cells in the nucleus, protected by the most 
external ring constituted by $A_1$ cells, will increase as $t^D$.  
Behind the first ring, the three species are equivalent, then the 
total number of cells occupied by each species in the nucleus is 
expected to increase with the same law $t^D$ too. 
In fact, this behavior is observed in numerical simulations, 
as shown in Fig.~\ref{fig:n1n2}, 
although species $A_1$ yields a slightly larger 
slope than $A_2$. Linear growth implies a regime where the concentrations 
in the nucleus are conserved. 
When full occupation of the lattice is attained, 
the concentration of each species $c_i=n_i/L^D$ ($i=1,\ldots,3$)  
fluctuates around the stable center predicted by the mean-field theory: 
$c_i=k_i/\sum_j k_j = 1/3$, for all $i$ \cite{pnb99}.

We have seen that the production  of the species in the  nucleus  is such that
after a short  transient their concentrations remain approximately  constant.
Thus, in the extreme case when windows have the size of a cell ($l=1$), 
a linear increase of $S_q$ with time occurs clearly for $q$ given by Eq. (\ref{qc}). 
In fact there will be $n_1$ windows with non-null probability $1/n_1$, where 
$n_1(t)\sim t^D$, then $S_q(t)=(n_1^{1-q}-1)/(1-q)$, 
so that: $S_q\sim t$, if $(1-q)D=1$. 
In the opposite extreme where windows are so large that all non-empty   
windows have approximately 
the same occupation number, that same temporal dependence is obtained too. 

\begin{figure}[htb]  
\begin{center}
\includegraphics*[bb=60 135 530 748, width=0.35\textwidth]{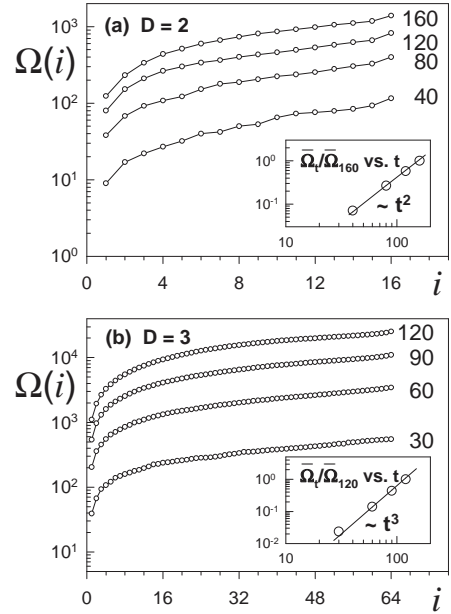} 
\end{center}
\caption{Cumulative distribution of occupation numbers at different 
times (in MCS) indicated in the figure for $D=2$ (a) and $D=3$ (b). 
$\Omega(i)$ is the number of windows occupied by less than $i$ cells 
of species $A_1$ ($1\le i\le l^D$).
In all cases, the window edges have length $l=4$.  
For $D=2$, $(L,\lambda)=(200,6)$,  and for $D=3$, $(L,\lambda)=(148,4)$. 
Insets: multiplication factor $\bar\Omega_t/\bar\Omega_{t_{max}}$ 
as a function of time (in MCS), 
where the horizontal bars mean average over all the occupation numbers 
at a given $t$,  and $t_{max}=$ 160 (a), 120 (b). 
}
\label{fig:occupation}
\end{figure}

Let us analyze what happens for intermediate window sizes. 
Clusters of cells of each species like in fully occupied 
lattices \cite{tpt04,fractals} appear in the interior of the nucleus. 
It is not counterintuitive the idea that  
as the volume of the nucleus is increasing, 
the size  distribution of the agglomerates becomes stationary 
after a transient.  
To test this idea one can calculate for instance the occupation numbers 
of the windows, intimately related to entropy computation, 
and see how their distribution evolves in time.  
Fig.~\ref{fig:occupation} exhibits the number of windows 
$\Omega(i)$ occupied by at most $i$ cells of species $A_1$ ($1\le i\le l^D$), 
for different time instants. The occupation number for $i=0$ is dismissed 
whereas it does not contribute to the entropy and a cumulative probability 
is considered in order to get smoother curves.  
Notice that the curves in the log-linear plot are practically parallel, 
indicating that they differ in a multiplicative factor. 
Moreover, the replication factor increases following the law $t^D$ as 
illustrated in the inset of Fig.~\ref{fig:occupation} and in  accord with 
previous considerations. 
All this means that a replication of the windows as described 
in Sect.~\ref{sec:cases} with  $\gamma=2$ is going on.

In conclusion, although the underlying dynamics is   
complex, once the  properties of spatial structures attain steady values, 
growth  is controlled by simple laws. 
More  specifically,  the resulting growth processes are power-law in time 
with scaling exponents trivially related to the lattice dimensionality   
(with $\sigma\sim t$ or, equivalently, $\gamma=2$). 
Then,  the linear increase of $S_q$ with time for $q_c$ 
given by Eq.~(\ref{qc}) is expected.

\section{Final remarks}

In this paper we employed the criterion of finite asymptotic growth 
rate of $S_q$ to determine $q_c$ in the $D$-dimensional LLV. 
We verified for $D=3$ and $D=4$ a conjecture for the connection between $q_c$ and $D$  
previously proposed \cite{tpt04}, namely, $q_c=1-1/D$. Moreover we exhibited the  
mechanisms leading to such relation. 

The box-counting method applied to the 2D LLV \cite{fractals} has shown that the 
boundaries of the small domains of a given species 
are approximately fractal \cite{fractals} with a fractal dimension $d_f$ that  
depends  on the reaction  rates $k_i$. 
Analogous  results are expected in higher dimensions. 
We have seen that, for the present model and for the particular probability 
assignment considered, $S_q$ entropy production does not capture directly 
the fractality of spatial patterns.  
Of course, since the fractal dimension $d_f$ must depend on  
the lattice spatial dimension $D$, the characteristic entropic index 
$q_c$ (a function of $D$) results in some way connected to $d_f$. 
But $q_c$ is not determined by the degree of fractality, 
it is only determined by $D$, 
independently of the nature (fractal or not) of the growing core.  
This is so because the properties of spatial patterns, such as the interface 
roughness, soon reach steady values. 

In a process where the number of occupied cells increases exponentially 
with time  $t$, the usual BGS entropy $S_1$ increases linearly with $t$. 
For a growth process that is not exponential, one can not expect 
a linear increase of  the standard  entropy $S_1$. 
If growth occurs following a power-law in time, $S_q$ has the property of 
finite production rate for some $q=q_c\neq 1$. 
Then $q_c\neq 1$ is the expected behavior in theses cases and it is not 
necessarily connected to the complex features of a system. 
So $q_c$ may result trivially related to  the lattice 
dimensionality.

We have shown that the stationarity of certain properties of the LLV dynamics determines 
a growth process linear in time ($\gamma=2$), yielding relation (\ref{qc}).  
Although simple,  this relation is in some way a consequence of the complex LLV dynamics. 
As perspectives, one can not exclude the  possibility  that 
in other dynamical regimes of the LLV, the increase of $S_q$  can 
reflect complex features directly. Also, 
it could be insightful to review previous works in the literature 
by taking into consideration the present results. This might be especially 
fruitful in cases where characteristic indexes of the form 
$q_c=1-2/(\gamma D)$ have also been found, as in the interesting study of 
Galilean-invariant lattice Boltzmann models of fluids \cite{boghosian}.

\section*{Acknowledgments:} 
I am very grateful to Fulvio Baldovin and Constantino Tsallis 
for interesting and fruitful discussions. 
I also thank Astero Provata for useful comments on the LLV. 
This work was supported by Brazilian agencies FAPERJ and PRONEX.

\end{multicols}


\begin{thebibliography}{99}
\bibitem{poison} Occasionally, if the initial droplet is too small, 
it may be invaded by one of the species.
\end{thebibliography}
\end{document}